\documentclass[11pt,a4paper]{article}
\usepackage{cite}
\usepackage{amsmath, amsthm, amssymb,slashed}
\usepackage{dsfont}
\usepackage{eufrak}
\usepackage{yfonts}

\usepackage{graphicx}

\usepackage{fancyhdr}
\usepackage{mathbbol}
\usepackage{mathabx}
\usepackage{url}
\usepackage{color}
\usepackage{bbm}
\usepackage{dsfont}
\usepackage{bm}
\usepackage{amsfonts}

\usepackage{datetime}

\def\be{\begin{equation}}
\def\ee{\end{equation}}
\def\bea{\begin{eqnarray}}
\def\eea{\end{eqnarray}}

\def\gev{\mathrm{GeV}}

\parskip=\baselineskip

\def\1{{\bf 1}}
\def\2{{\bf 2}}
\def\3{{\bf 3}}
\def\4{{\bf 4}}

\def\hat{\widehat}
\font\teneurm=eurm10 \font\seveneurm=eurm7 \font\fiveeurm=eurm5
\newfam\eurmfam
\textfont\eurmfam=\teneurm \scriptfont\eurmfam=\seveneurm
\scriptscriptfont\eurmfam=\fiveeurm

\font\teneusm=eusm10 \font\seveneusm=eusm7 \font\fiveeusm=eusm5
\newfam\eusmfam
\textfont\eusmfam=\teneusm \scriptfont\eusmfam=\seveneusm
\scriptscriptfont\eusmfam=\fiveeusm

\font\tencmmib=cmmib10 \skewchar\tencmmib='177
\font\sevencmmib=cmmib7 \skewchar\sevencmmib='177
\font\fivecmmib=cmmib5 \skewchar\fivecmmib='177
\newfam\cmmibfam
\textfont\cmmibfam=\tencmmib \scriptfont\cmmibfam=\sevencmmib
\scriptscriptfont\cmmibfam=\fivecmmib

\numberwithin{equation}{section}

\def\c{{\mathcal C}}
\def\en{{\mathcal E}}
\def\n{{\mathcal N}}

\begin{document}
\begin{titlepage}
\begin{flushright}

\end{flushright}
\vskip 1.5in
\begin{center}
{\bf\Large{Directional Dark Matter Searches with Carbon Nanotubes}}
\vskip 0.9cm {LM Capparelli$^*$, G Cavoto$^\P$, 
D Mazzilli$^*$ and AD Polosa$^{*,\P}$} \vskip 0.05in {\small{ \textit{$^*$Dipartimento di Fisica, Sapienza Universit\`a di Roma, Piazzale Aldo Moro 2, I-00185 Roma, Italy}\vskip -.4cm
{\textit{$^\P$INFN Sezione di Roma, Piazzale Aldo Moro 2, I-00185 Roma, Italy }}}
}
\end{center}
\begin{center}
\makeatletter
\tiny\@date
\makeatother
\end{center}
\begin{abstract}
A new solution to the problem of dark matter directional detection might come from the use of
large  arrays of aligned carbon nanotubes. 
We calculate the expected rate of carbon ions channeled in  single-wall nanotubes 
once extracted by the scattering with a massive dark matter particle. Depending on its initial kinematic conditions,  the ejected Carbon ion  may be channeled in the nanotube array or stop    in the bulk.  The orientation of the array with respect to the direction of motion of the Sun  has an appreciable effect on the channeling probability. This provides the required  anisotropic response for a directional detector. 
\end{abstract}
%
\end{titlepage}

\section{Introduction}
The only striking evidence of the  Weakly Interacting Massive Particle (WIMP) nature of  dark matter reported so far comes from the annual modulation in nuclear recoil rates observed by the DAMA/LIBRA experiment at INFN Gran Sasso, Italy~\cite{dali}.
Their impressive results have not found any confirmation by other direct dark matter search experiments although we can mention some interesting indications by  CoGENT~\cite{cogent}, CDMS(Si)~\cite{cdms} and  CRESST~\cite{cresst}. 
There are important experimental results~\cite{Akerib:2013tjd}, obtained with different techniques, in clear contradiction with  the DAMA/LIBRA ones. 

However light WIMPs ($\lesssim 10$~GeV) are still worth studying as possible constituents of a diffuse dark matter galactic halo~\cite{spergel}. 
The rare interactions of WIMPs with detector nuclei (sodium or iodine nuclei, or even electrons~\cite{damaele}, as in the case of DAMA/LIBRA) might cause the latter  to recoil out of their crystal sites releasing energy through ionization. Yet nothing can be concluded, with the apparatuses being used so far, about their (average) direction, namely the WIMP wind direction. 

In other words, dark matter direct search experiments are presently unable to discern nuclear recoil directions.
Among the proposals to solve this problem we quote for example  the one by the DRIFT collaboration exploiting low mass gas detector~\cite{drift},  the use of  nuclear emulsions~\cite{nuclemul}, the use of DNA molecules as  WIMP targets~\cite{Drukier:2012hj}.
 
In this paper we address the issue of directional detection of dark matter with a new approach: carbon ions channeling within  single-wall carbon nanotubes (CNT). The unit component of the ideal detector we want to study is a CNT closed at one end and open at the other. Its body provides at the same time the {\it target} for WIMP-carbon nuclei collisions and the {\it channel} to direct the torn out nuclei towards the free end.

We present a first study of this kind of directional detector through the numerical simulation of WIMP collisions on the surface of CNTs and the subsequent channeling of  recoiling carbon nuclei eventually escaping the CNTs at their open ends. Our simulations are based on the calculation of the rate of nuclear recoils due to WIMP-nuclei interactions as a function of the nuclear final kinetic energy $T$ and  recoil angle $\theta$ with respect to the WIMP wind direction. 

We find that a single CNT  may serve as an Êanisotropic and direction discriminating device.
Different orientations of a CNT array with respect to the WIMP wind, give sensibly different channeling probabilities and different ion counts at their free ends.  
The main point is that  CNTs are very efficient at channeling since they are rather empty channels: contrary to what happens in crystals, ions traveling through carbon nanotubes have very little interactions with electrons. This  drastically decreases the {\it dechanneling} probability.

Arrays of carbon nanotubes allow a secondary channeling which strengthens the directional signal we are seeking. Interstices among nanotubes are themselves contributing to give an overall channeling effect, once the right scattering initial conditions are met. 

\section{WIMP-Carbon scatterings}
\begin{figure}[ht]
 \begin{center}
   \includegraphics[width=9truecm]{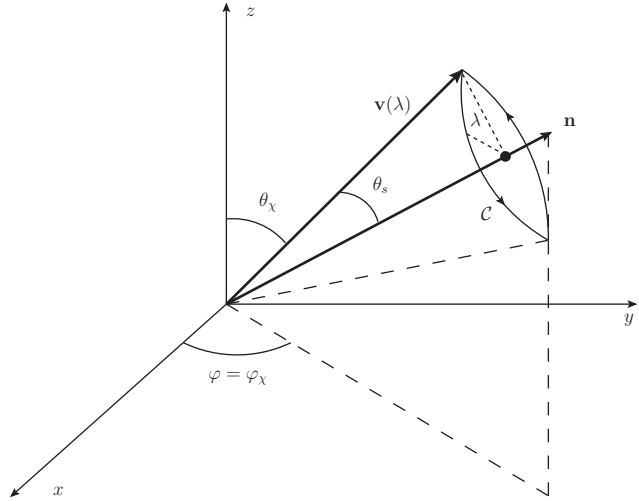}
 \end{center}
\caption{\small The $z$ axis is in the direction of $\bm w$ given in~(\ref{wut}). 
The angle $\theta_\chi$ is relative to the incoming dark matter (DM) particle $\chi$. The angles $\theta (=\theta_\chi+\theta_s~\text{in figure})$ and $\varphi$ define the direction of the recoiling nucleus $\mathcal N$, along $\bm n$, in the elastic  scattering process $\chi+\mathcal N\to \chi +\mathcal N$. The incoming DM particle direction is unknown but, if we fix the nucleus recoiling (kinetic) energy $T$ and the incoming $\chi$ velocity $|\bm v|$, the  cone semi-aperture $\theta_s$ gets fixed.  Given $T$ and $|\bm v|$, the $\chi$ initial velocity vector tip will lie somewhere on  $\mathcal C$. Averaging out the initial $\chi$ velocity direction translates into integrating the cross section, with some velocity distribution $f(\bm v)$, along $\c$, for given values of  $\theta_s$.   \label{frame}}
\end{figure}

Let $\bm v$ be the velocity of the incident dark matter (DM) particle $\chi$ and $\bm n$ be the direction ($|\bm n|=1$) of the scattered nucleus $\n$. Define the frame in Fig.~\ref{frame} with the $z$-axis along the Sun's direction. In this frame the velocity vector  
\be
\bm w(t)\simeq \left[232+15\cos\psi(t)\right]{\bm k}
\label{wut}
\ee
is oriented along the Sun's direction and its length takes into account the velocity of the Earth around the Sun
\be
\psi(t)=\left(2\pi \frac{t-152.5}{365.25}\right)
\label{psit}
\ee
The $\chi$ `wind' vector experienced by a detector at the origin of this frame will be $-\bm w$.
The velocities are expressed in km/sec and the time is in days starting from January 1st.  Here $232$~km/sec is the Sun's velocity along its Galactic motion. The value of 15 km/sec is approximately equal to $30~\text{km/sec}\times \cos 60^\circ$, $30$~km/sec being approximately the velocity of the Earth along its motion around the Sun and $60^\circ$ the inclination of the plane of the Earth's orbit with respect to the Sun's one.

We obtain the formula
\be
\frac{d\Gamma}{dT\, d\cos\theta}=K\frac{n_\chi G^2_A\, \Lambda^4\,F_A(2M_\n T)^2}{16 M_\chi^2  M_\n}\int_{\sqrt{2T/\mu}}^{v_{\rm max}} dv\,v\, \beta\, e^{-\alpha\, A(v,T,\theta,t)}\, I_0(\alpha \,B(v,T,\theta,t))
\label{rate2}
\ee 
where $n_\chi$ is the incident dark matter (DM) number density, $G_A$ is some unknown effective DM-nucleus interaction coupling, $F_{A}(\bm p^{\prime 2})$ is (adimensional) nuclear form factor and the scale $\Lambda$ has GeV dimensions and sets the units -- both $G_A$ and $F_A$ depend on the  $A$ nucleus involved. $I_0$ is the modified Bessel function of the first kind. 
In numerical calculations we will assume $\Lambda\sim \sqrt{s}\simeq M_\chi+M_\n$. We are especially interested in the maximum-energy trasfer region which is found  at  $M_\chi\approx M_\n$.

In order to express $d\Gamma/dTd\cos\theta$ in units of counts-per-day, per kg, per keV (cpd/kg/keV), formula~(\ref{rate2}) needs to be multiplied by $K=5\times 10^7$ -- this includes also the number of target nuclei per kg in carbon nanotubes. With this value of $K$, velocities in the integral must be expressed in units of $c$. Otherwise, km/sec can be used but with  an overall coefficient  $K=1.5\times 10^{13}$.

The maximum velocity is $v_{\rm max}=232+550$~km/sec~\footnote{The escape velocity at the Solar System's galactic radius with respect to the Milky Way's gravity is $\approx 550$~km/sec. The Maxwell-Boltzmann  velocity distribution in the Galactic frame is sharply  limited at $v_{\rm max}$ by a factor $\theta(v_{\rm max}-V)$. In the calculation of the rate~(\ref{rate2}) we have $V=v+w$ so that $v_{\rm max} - w =v_{\rm esc}= 550$~km thus allowing $v_{\rm max}\approx 780$~km/sec. }  
whereas 
\be
\mu=\frac{4M_\chi^2M_\n}{(M_\chi+M_\n)^2}
\ee
and $T$ is the nucleus recoiling kinetic energy (in the non-relativistic limit we use). The definition of the functions $A$ and $B$ in the integrand are 
\bea
\label{uno}
A(v,T,\theta,t)&=&v^2+w^2(t)+2v\cos\theta_s \cos\theta\, w_z(t)\\
\label{due}
B(v,T,\theta,t)&=&2v\sin\theta_s\sin\theta\,w_z(t)
\eea
where $\bm w$ is given in~(\ref{wut}). Observe that there is no dependency left on the  $\varphi$ angle and 
\be
\cos\theta_s=\sqrt{\frac{2T}{\mu v^2}}
\ee
The $\alpha$ and $\beta$ parameters characterize the WIMP velociy distribution defined by
\be
f(\bm v+\bm w)=\beta e^{-\alpha (\bm v+\bm w)^2}
\ee
where $\alpha\sim 1/v_0^2,\beta\sim 1/v_0^3$ ($v_0$ is usually taken to be $v_0\simeq235$~km/sec)\footnote{The (adimensional) numerical values for $\alpha$ and $\beta$ in our calculation are $\alpha=1.6\times 10^6$ and $\beta=3.7\times 10^8$.}, and the distribution peaks at $\bm v=-\bm w$, {\it i.e.} along the $\chi$ particles wind, see Eq.~(\ref{wut}). As for $n_\chi$ we use $n_\chi=\rho_\chi/M_\chi=(0.4\pm0.04~\mathrm{GeV}/\mathrm{cm}^3)/11~\mathrm{GeV}\approx 0.04~\mathrm{cm}^{-3}$ considering WIMPs with mass close to the carbon mass. 

As for the $G_A$ value for WIMP-Carbon scatterings to be used in formula~(\ref{rate2}) we compute in the Appendix the value
\be
G_A= 1.6\times 10^{-5}~\mathrm{GeV}^{-2}
\ee 
with $M_\chi\simeq M_\n$. This value is extracted from a comparison with DAMA data on modulation for WIMP-Na scatterings for
WIMPs of mass $M_\chi\simeq 11~\gev$. It corresponds to WIMP-nucleon cross sections of $\sigma_{\chi p}=2\times10^{-4}$~pb.

The non-relativistic kinematics of the DM-nucleus scattering involves the factorization of the nuclear form factor so that the angular distribution is universal, depending only on the Maxwell-Boltzmann assumption of the DM velocity distribution.  Carbon is a spin zero nucleus, therefore our analysis concerns spin independent interactions. 

As for the form factor, following the standard assumptions made in the literature~\cite{lis}, we use the Helm function
\be
F_A(q)=3\,e^{-q^2s^2/2}\,\frac{\sin(qr_n)-qr_n\cos(qr_n)}{(qr_n)^3}
\ee
where $s\simeq0.9$~fm and $r_n^2=c^2+(7/3)\pi^2a^2-5s^2$ is an effective nuclear radius  with $a\simeq 0.52$~fm and $c\simeq 1.23\,A^{1/3}-0.60$ in units of fermis; $q$ is given by $q=\sqrt{2m_\n T}$. 

\begin{figure}[htb!]
 \begin{minipage}[c]{8cm}
   \centering
   \includegraphics[width=6.5cm]{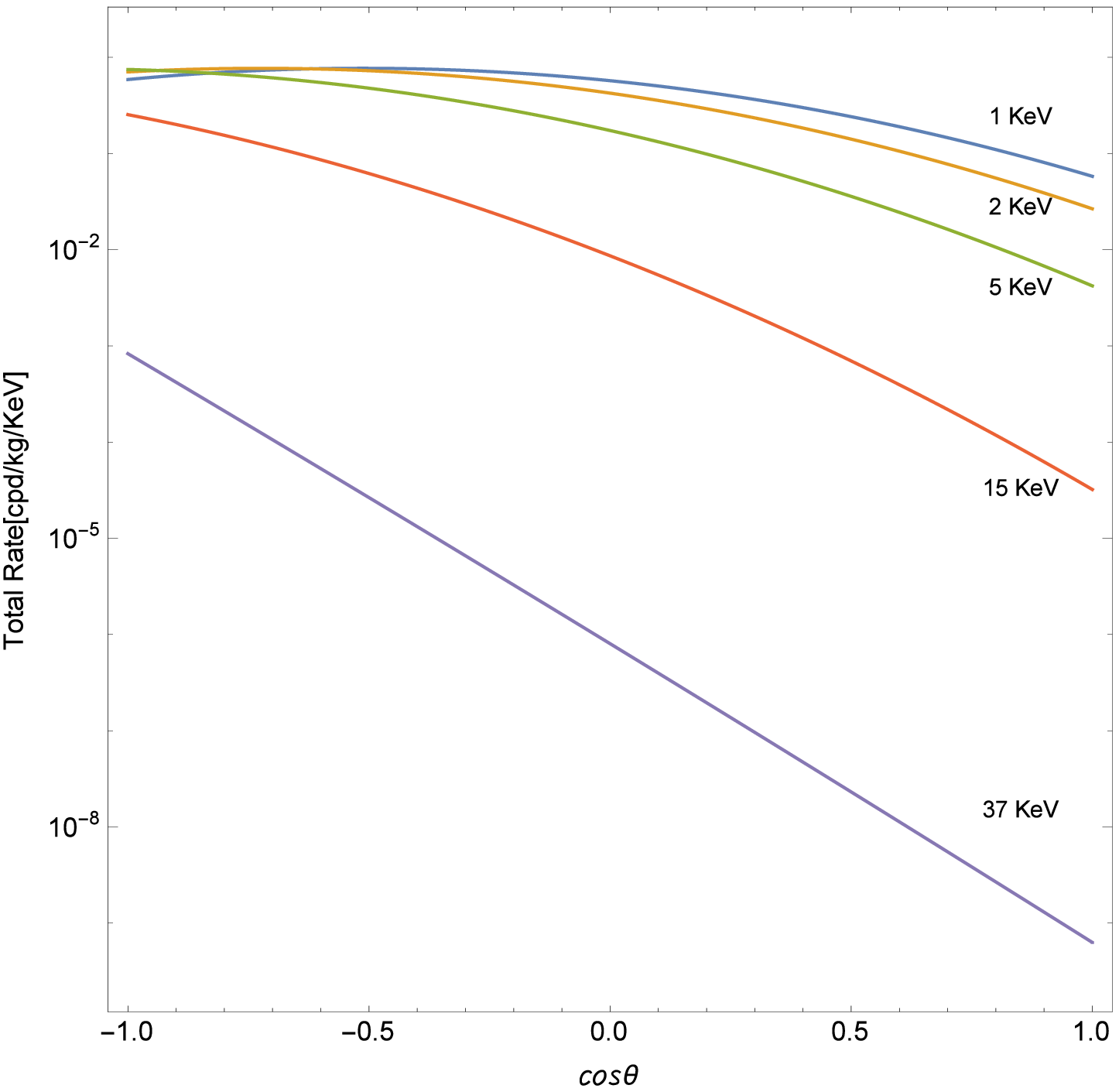}
 \end{minipage}%
 \begin{minipage}[c]{5cm}
\centering
   \includegraphics[width=6.5cm]{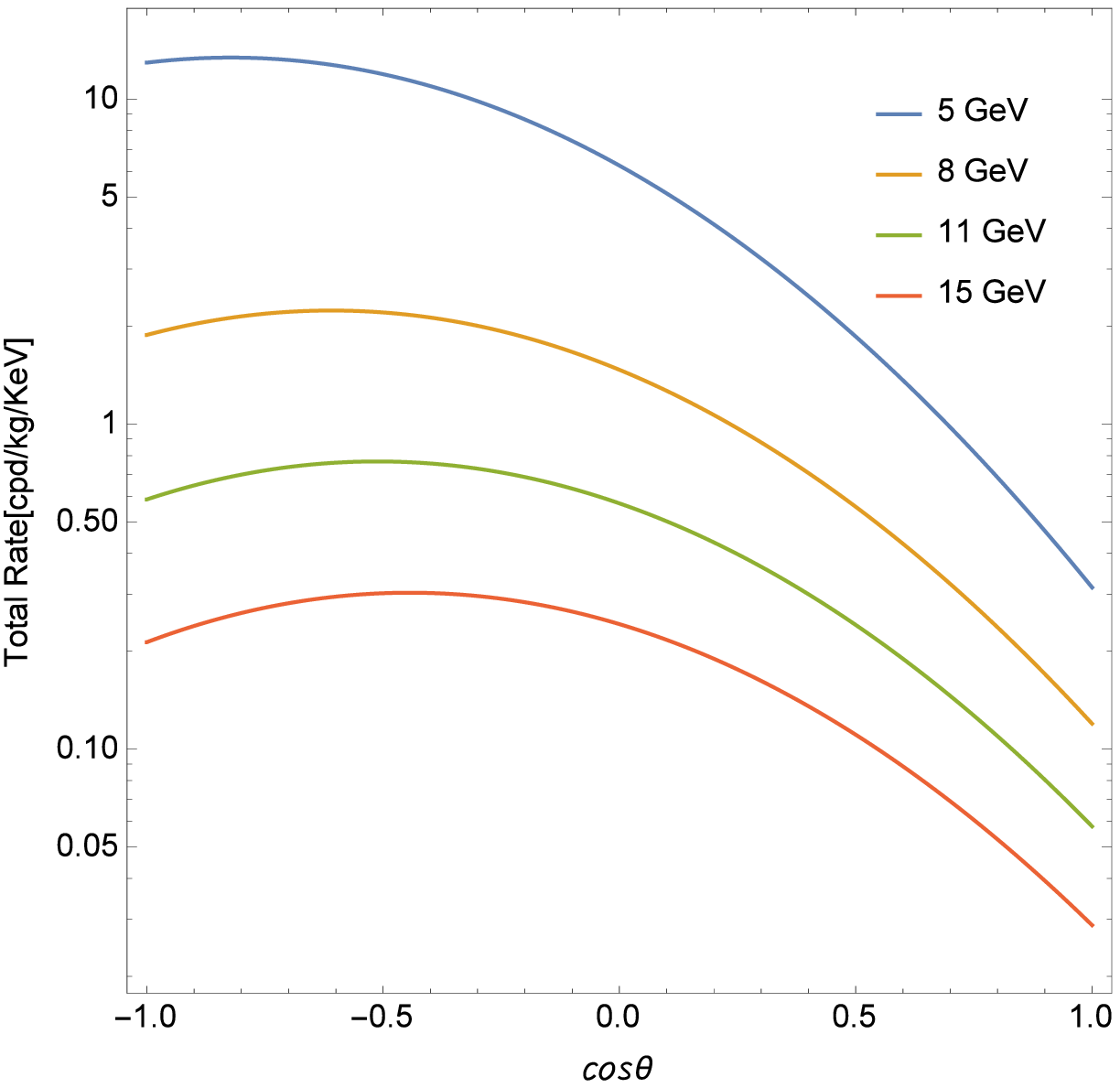}
 \end{minipage}%
 \caption{\small Left Panel: Angular distributions in the $\cos\theta$ variable, $\theta$ being the angle of the recoiling nucleus with respect to the Sun direction at the scattering time. Curves for various values  of $T$ are displayed. Here $M_\chi=11$~GeV and $\sigma_{\chi p}\approx10^{-4}~\mathrm{pb}$ is used. Right Panel: Same as left panel  for different values of $M_\chi$ with recoil kinetic energy $T=1$~keV (the minimum allowed value in our analysis). 
 The variation with time is found to be negligible. The figures are related to carbon nuclei. 
 \label{angdistrc}}
\end{figure}

We will use the distribution $d\Gamma/dT d\cos\theta$ in~(\ref{rate2}) -- see Fig.~\ref{angdistrc} for the case of carbon nuclei and $M_\chi\approx M_\n$ --  as the probability distribution to randomly generate events of recoiled nuclei carrying kinetic energy $T$ with a scattering angle  $\theta$ with respect to the WIMP wind direction. Namely we generate a sample of $(T,\theta)$ pairs in the low mass window $M_\chi\approx 11$~GeV. This represents a set of possible initial conditions of the Carbon ion in the nanotube. In the next section we will attribute a 
weight $\mathbb{w}(T,\theta)$ to the $(T,\theta)$ event which corresponds to   the probability that initial conditions meet channeling conditions in nanotubes.


\section{Positive ion motion in a CNT}
\begin{figure}[htb]
   \centering
   \includegraphics[width=6cm]{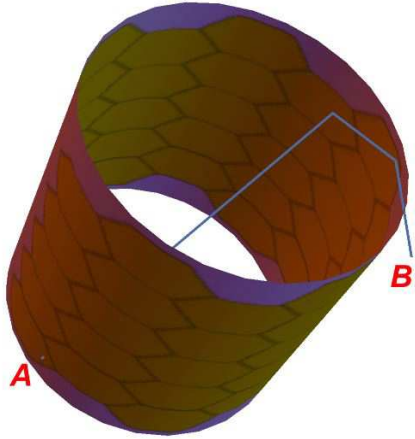}
 \caption{\small The WIMP particle hits a carbon nucleus of the nanotube in $A$. The carbon nucleus can recoil within the nanotube and have the correct parameters to be totally reflected within the CNT until it escapes in $B$.   \label{nanot}}
\end{figure}

As shown in Fig.~\ref{nanot}, a WIMP collides with the surface of a carbon nanotube  in point $A$, extracting a carbon nucleus from it. This might be kicked inside the nanotube and, provided  it has the correct parameters in terms of kinetic energy $T$ and direction $\theta$ (with respect to the CNT axis), it will be {\it channeled} in the CNT, a phenomenon similar to total reflection in an optical fiber. The carbon nucleus eventually reaches one of the two ends of the CNT. Actually, because of {\it crystal blocking}, to be discussed below, the carbon nucleus has a chance of being channeled only if it is struck by the WIMP while its position fluctuates inward due to thermal vibrations.  

On the basis of the recoil distributions $d\Gamma/dT d\cos\theta$, we would roughly expect that a CNT oriented along the WIMP wind direction ($z$ axis in Fig.~\ref{frame}) will channel more effectively the hit carbon nuclei  with respect to the case in which the nanotube is oriented perpendicularly to the WIMP wind, see Fig.~\ref{angdistrc}.
Once the channeled carbon ion has escaped the CNT, it is collected and registered by a gas detector (point $B$ in Fig.~\ref{nanot}).

The Carbon nanotube injected in the CNT scatters coherently with all atoms of the tube and so experiences an effective axial potential (energy) along its transverse motion (with respect to the CNT axis) given by~\cite{russi} 
\be
U(r,\phi)=U_0(r)+2\sum_{s=1}^{\infty} U_{sN}(r)\cos\frac{\pi s(n+m)}{q}\cos\left(sN\phi+\frac{\pi s(n+m)}{q}\right)
\label{axial}
\ee
where cylindrical coordinates $(r,\phi)$ are used.
The $U_\nu (r)$ terms, $\nu=0,1,2...$, are independent of the $\phi$ angle taken from the center of the potential -- they have a cylindrical symmetry.  It is found that  
\be
U_{\nu}(r)=4\sqrt{\pi}\sigma Z^2 e^2\left(\frac{R}{r}\right)^{1/2}\sum_{j=1}^4 a_jb_j\, e^{-b_j^2(r^2+R^2)} e^{2 b_j^2 rR\left[\sqrt{1+\xi^2}-\xi\ln (\xi+\sqrt{1+\xi^2})\right]}
\label{axial2}
\ee
where $\xi=\nu /(2b_j^2 r R)$ and Gaussian units are used ($\alpha=e^2/\hslash c$); $r$ assumes values from 0, the center of the potential, to the surface of the CNT at $r=R$. Here $\sigma=4/(\ell^2\sqrt{27})$ where $\ell\simeq 0.14$~nm is the bond length between carbon atoms and
\bea
&&a_j=(3.222,5.270,2.012,0.59499)\times 10^{-4}~{\rm nm}^2\notag\\
&&b_j=(10.33,18.694,37.456,106.88)~{\rm nm}^{-1}
\eea
as reported in~\cite{russi}. The exponential function in the sum in~(\ref{axial2}) reduces to $e^{-b_j^2(R-r)^2}$ when $\nu=0$. 
The CNT are characterized by a {\it roll-up} vector $\bm r_0=n\bm a+m\bm b$, where $n,m$ are two integers, the CNT indices, and $\bm a,\bm b$ are the basis vectors of a graphite plane (the angle between $\bm a$ and $\bm b$ is $\pi/3$ and $|\bm a|=|\bm b|=\ell\sqrt{3}$). 
The nanotube is a rolled up strip of width $|\bm r_0|$ cut out by a graphite carbon plane 
perpendicularly to $\bm r_0$. The resulting cylinder is closed (at either end) with caps containing carbon penthagons in a manner that conserves bonding length. 
The radius $R$ of a CNT is 
\be
R=\frac{\ell\sqrt{3}}{2\pi}\sqrt{n^2+nm+m^2}
\ee
A nanotube $(n,m)$ can be represented as a collection of $2N$ rows parallel to the tube axis and it is found that
\be
N=\frac{2}{q}(n^2+nm+m^2)
\ee
where $q={\rm gcd}(2m+n,2n+m)$. These equations fully define Eq.~(\ref{axial}).
\begin{figure}[htb!]
   \centering
   \includegraphics[width=6.5cm]{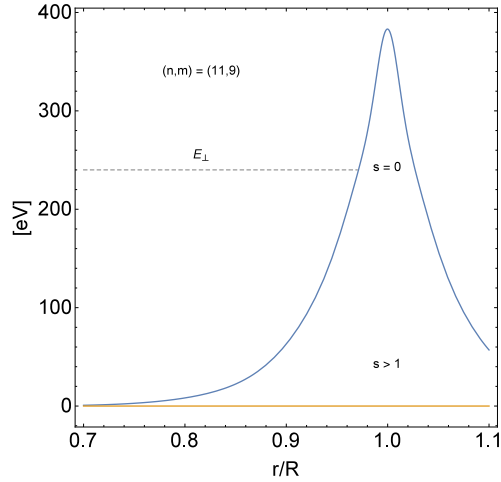}
 \caption{\small $(11,9)$ CNT. The profile of $U_0$, $(s=0)$, versus that of $U_{602}$, $(s=1)$. The effective potential $U(r,\phi)=U(r)$.
\label{axialpot}}
\end{figure}

The essence of the channeling phenomenon is the existence of a classical turning point as in Fig.~\ref{axialpot}. In other words the total transverse energy $E_\perp$ of the ion kicked inside the CNT by the WIMP, should not exceed the effective potential  barrier at $r\approx R$ for the ion to be channeled. In the Lindhard theory of channeling another condition is required~\cite{lind}: if $x$ is the distance from the nanotube surface, 
\be
U^{\prime\prime}(x)<\frac{8T}{(\ell \cos(\pi/3))^2}
\ee
which we find is verified for every $x$ in the range of kinetic energies $T$ we consider. 

\begin{figure}[htb]
   \centering
   \includegraphics[width=8cm]{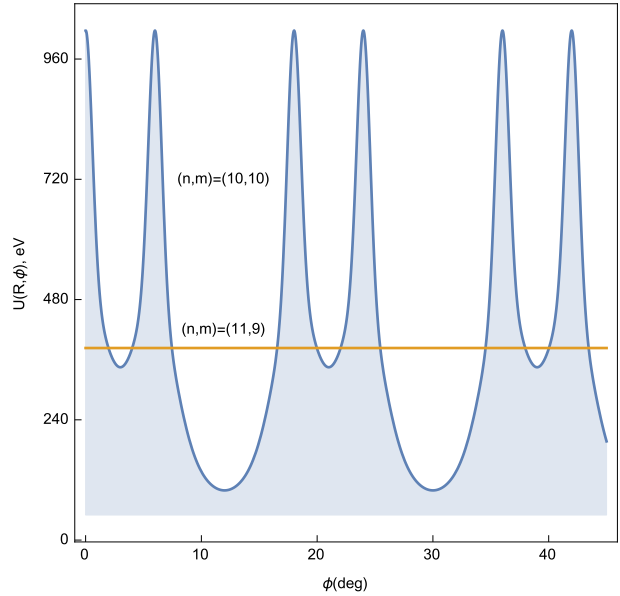}
 \caption{\small The $\phi$ dependence of the effective potential $U(R,\phi)$ at the nanotube surface $r\simeq R$ in the case of the $(10,10)$ CNT as compared to the $(11,9)$ one.  The $(10,10)$ nanotube clearly shows some `holes' with respect to the $(11,9)$ one (see the region around $\phi\sim 10^\circ$ or $\phi\sim 30^\circ$)  with the consequence of being less effective at channeling. \label{window}}
\end{figure}

In Fig.~\ref{axialpot} we show how for certain CNT, for example the $(11,9)$,
the higher $U_{\nu>0}$ terms are irrelevant and the effective potential has cylindrical symmetry.  In other cases, when $n=m$ or $m\vee n=0$, the higher terms are important and $U(r\approx R,\phi)$ of Eq.~(\ref{axial}) has an explicit dependence on $\phi$, showing a periodical sequence of maxima and minima, see Fig.~\ref{window}.  

Therefore, besides some particular cases, the motion of the ion in the transverse direction is subject to a potential $U(r,\phi)$ with a sinusoidal $\phi$ dependency. This means that the $v_\phi=r\dot{\phi} $ component of the carbon ion transverse velocity will change when approaching $r\approx R$, and the transverse angular momentum will not be an integral of motion. On the other hand, for CNT like the  $(11,9)$ of Fig.~\ref{window}, a centrifugal barrier prevents channeled trajectories (those having $E_\perp$ lower than the barrier at $R$) to get very close to the CNT axis.

The temperature of the system, $T^\star$, might also be taken into account when deriving the effective potential $U(r,\phi)$. However, it is shown that the influence of thermal vibrations is relatively small in the definition of function $U(r,\phi)$. 

On the other hand, at finite $T^\star$, each single carbon site undergoes transverse thermal fluctuations (longitudinal ones are irrelevant here), bringing it toward the inside (or the outside) of the CNT. The one-dimensional amplitude of thermal fluctuations in Debye theory is given by $u_1(T^\star)\approx 10^{-2}$~nm at room temperature. As discussed in the context of crystals~\cite{gelmi},  due to lattice vibrations the collision with a WIMP may happen while a carbon site, in our case, is far enough within the interior of the channel. This is essential to channeling:  if channeled ions never go close to lattice sites the reversed paths do not happen either. 

\begin{figure}[htb]
   \centering
   \includegraphics[width=6cm]{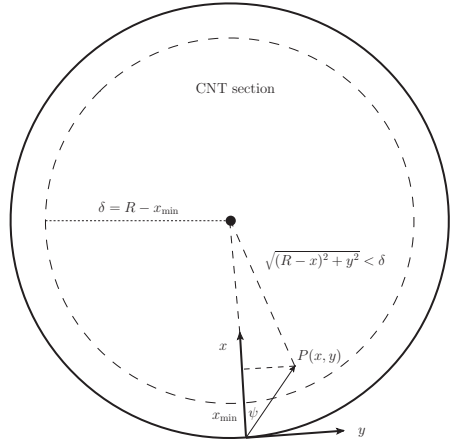}
 \caption{\small Due to lattice vibrations, a carbon site fluctuates from the CNT surface to point $P$. For channeling to take place, point $P$ has to be inside a disk of radius $R-x_{\rm min}$. The WIMP-carbon scattering is supposed to occur at point $P$.  \label{mincond}}
\end{figure}

To check the channeling conditions starting from a $(T,\theta)$ nucleus recoil event, we compute the transverse kinetic energy $T(\sin\theta)^2$ and require that the total transverse energy
\be
E_\perp=T(\sin\theta)^2+U(R-x,\phi)\lesssim \min U(R,\phi)
\label{eperp}
\ee 
This provides  a $x_{\rm min}$ minimum value for $x$.
The probability (weight of the event) that the carbon nucleon which recoiled with $\approx T\theta^2$ gets channeled is given by
\be
\mathbb{w}(T,\theta)=\int_{\cal R} dxdy\, \frac{e^{-x^2/2u_\perp^2(T^\star)}}{\sqrt{2\pi}u_\perp(T^\star)}\, \frac{e^{-y^2/2u_\parallel^2(T^\star)}}{\sqrt{2\pi}u_\parallel(T^\star)}
\label{weight}
\ee
where, see Fig.~\ref{mincond} 
\be
{\cal R}:=(R-x)^2+y^2<(R-x_{\mathrm{min}})^2
\ee
whereas the average radial and longitudinal vibration amplitudes (as described in Debye theory) are given by
\bea
&& u_\perp=0.0085~{\rm nm}\notag\\
&& u_\parallel=0.0035~{\rm nm}
\eea
\begin{figure}[htb!]
   \centering
   \includegraphics[width=6.5cm]{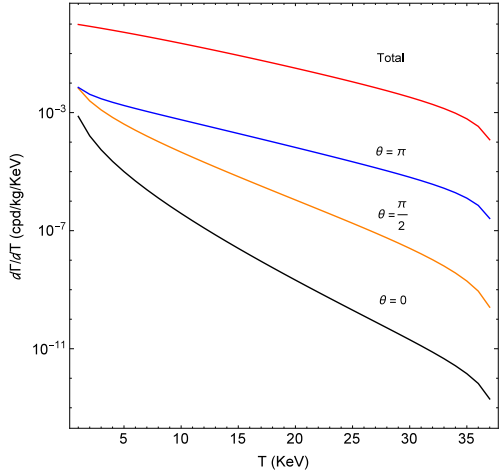}
  \caption{\small Here we report the main outcome of our numerical analysis: the carbon nanotube anisotropic response. Expected histogram of counts of channeled carbon nuclei in a CNT (with a diameter $d=10$~nm) as a function of the nucleus recoil kinetic energy $T$. It takes into account the predicted channeling efficiency and   $\sigma_{\chi p}\simeq 10^{-4}$~pb ($10^{-40}$~cm$^2$). The CNTs are oriented with the direction from the base to the open cap  along  $- z$, indicated as $\theta=\pi$, orthogonal to it, $\theta=\pi/2$,  and with $\theta=0$.  This specific plot has been obtained for  a $(n,m) = (73,71)$ nanotube.
   \label{channelings}}
\end{figure}
We therefore obtain the number of counts per day per Kilogram of channeled carbon nuclei expected in a CNT  using $i)$ the distribution~(\ref{rate2}) 
and $ii)$ the Monte Carlo calculation outlined in~(\ref{eperp}) and~(\ref{weight}). Namely we produce samples of $(T,\theta)$ events generated randomly as prescribed by the 
tuned distribution~(\ref{rate2}). To each of them we  attach the probabilistic weight $\mathbb{w}(T,\theta)$ and finally compute the histogram of channeled events as in~Fig.~\ref{channelings} using three orientations for the CNT: $\theta=\pi$, base to open cap of the CNT in the WIMP wind direction ($-\bm w$ of Eq.~(\ref{wut})), $\theta=\pi/2$, orthogonal to it and $\theta=0$, base to open cap opposite to the WIMP wind. 

In~Fig.~\ref{channelings} the CNT is supposed to be oriented {\it along} the $z$-axis of Fig.~\ref{frame}.  
\begin{figure}[htb!]
   \centering
   \includegraphics[width=8cm]{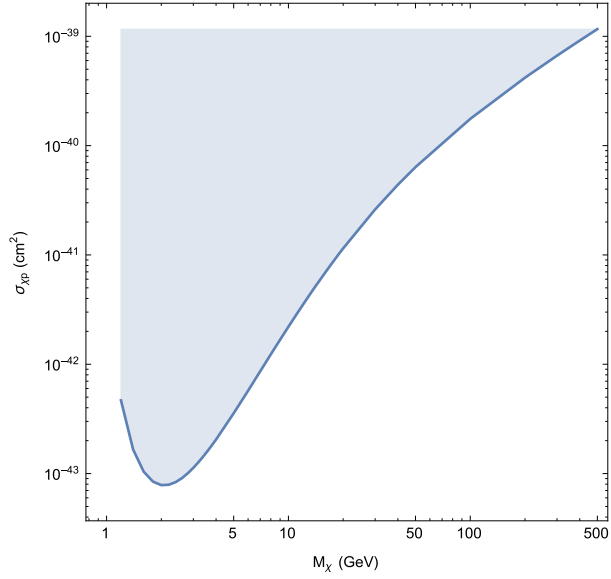}
 \caption{\small Exclusion region at 90$\%$ confidence level in the plane $\sigma_{\chi p}$~vs~$M_\chi$ for a 10~kg CNT target mass after 3 years exposure time, under the assumption that the  proposed detector  identifies  and rejects  all  background events and no signal events are recorded.  The behaviour at low mass is due to the energy of the scattered $C$ ion going below thekeV  1 keV threshold. \label{sensi}}
\end{figure}
In Fig.~\ref{sensi} we report the 
exclusion region at 90$\%$ confidence level in the plane $\sigma_{\chi p}$~vs~$M_\chi$ for a 10~kg CNT target mass after 3 years exposure time. This is obtained in the case of  no signal events being recorded and assuming a perfect discrimination of the backgrounds. 
We remark here that this sensitivity  is attainable  under the assumption of being able to detect C ion recoils down to 1 keV kinetic energy. 

\section{Numerical analysis of CNT arrays}
\begin{figure}[htb!]
   \centering
   \includegraphics[width=6.7cm]{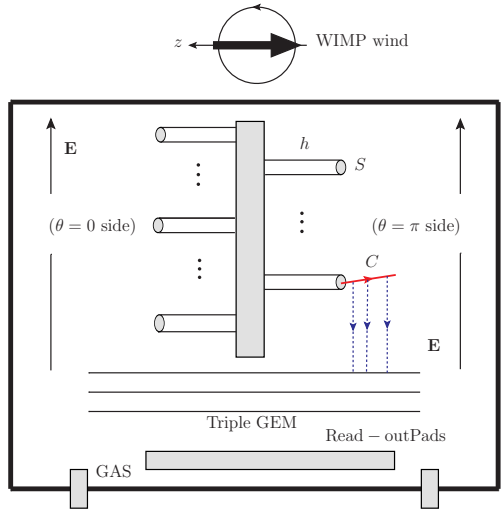}
 \caption{\small Bidimensional sketch of the CNT double array detector.  Nanotubes have a height $h$ which is typically $100\div 200~\mu$m and a section $S$ whose diameter is $\approx 10$~nm. A triple-GEM detector allows to collect the ionisation electrons  produced in the gas volume by a $C $ channeled ion (schematically represented in figure), thanks to an electric field orthogonal to the GEM plates. The substrate of the CNT arrays has a thickness of $\sim 1$~mm and a surface of $10\times 10$~cm$^2$. 
 \label{detector}}
\end{figure}

 As illustrated in the previous Section, a  single CNT exhibits an {\it anisotropic response} to 
 the interaction of heavy dark matter particles with carbon nuclei. We consider  the elementary detector component as composed by a  double  brush  of aligned CNTs attached to a substrate (Si -- for example), see Fig.~\ref{detector}. The two sides have the $\theta=0$ and $\theta=\pi$ response described in Fig.~\ref{channelings}. The full detector is supposed to be a large array of these modules. 

CNT brush configurations can easily be synthesized with a good alignment quality  on  various substrates.

In our conceptual design the substrate should be thick enough to absorb scattered ions directed towards it.  This CNT double-brush might eventually be inserted in a  time projection gas chamber  (TPC) to detect the channeled C ion, by measuring its kinetic energy and direction.  Other detection possibilities of the escaping carbon ion are left open. Indeed there are a number of technological solutions to be discussed elsewhere. The one reported here in Fig.~\ref{detector} is just for the sake of illustration. 

The CNT array in Fig.~\ref{detector} is placed in an electric field orthogonal to CNTs axes. This could perturb the structure of carbon atoms on the nanotube walls and therefore the shape of the effective potential $U(r)$. The magnitude of the perturbation $U'$ can be estimated to be the electric potential difference over the nanotube radius (the average circular segment is $\simeq 1.3 R$). A typical drift field is $|\bm E| \simeq 0.6 $~kV/cm producing a potential drop over a radius $R=5$~nm of $\Delta V \simeq 0.4$~mV. This allows to estimate $|U'|\approx Z e\Delta V \simeq 2.4$~meV for a C nucleus. The fact that $|U'|\ll |U|$ suggests that the external drift electric  field we would need does not  affect sensibly the channeling numerical analysis we are presenting.

An alternative solution might be that of integrating the carbon nanotubes layers as electrodes in a detection scheme as the one implemented in the DRIFT detector~\cite{Burgos:2007tt}, in which the electric field would be parallel to the CNT axes.

\begin{figure}[htb!]
   \centering
   \includegraphics[width=8cm]{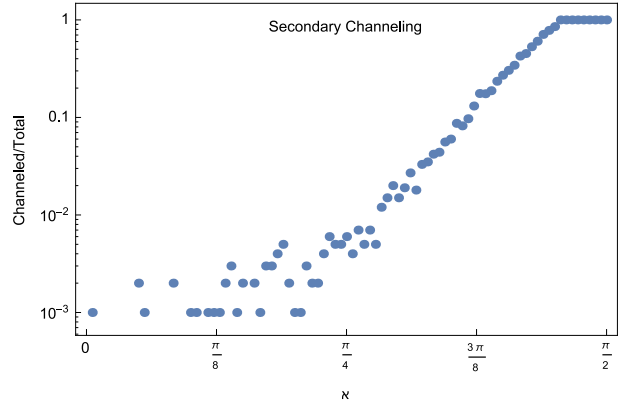}
 \caption{\small Fraction of secondary channelings obtained changing the angle $\aleph$ formed by the initial carbon ion direction with respect to the plane orthogonal to CNT axes. When $\aleph$ approaches $\pi/2$, secondary channelings are strongly enhanced but at this point they are any longer part of the background but a strong reinforcement of the signal.\label{altra2}}
\end{figure}

The outcome of our numerical simulations can be summarized as follows
\begin{enumerate}
\item {\bf \emph{Channeling conditions.}} Given the appropriate initial conditions in terms of recoil kinetic energy $T$ and scattering angle $\theta$ with respect to the single CNT axis $(T,\theta)$, the carbon ion may  or may not be channeled: the crucial point is that the transverse kinetic energy $E_\perp=T\theta^2$, for small $\theta$, is small enough when compared with the potential barrier experienced by the positive carbon ion when approaching the nanotube surface ($\sim O(400~\mathrm{eV})$).

\item {\bf \emph{Dechanneling along the nanotubes.}} A unitary cell of the nanotube contains two carbon atoms. Six out of the 8 electrons ($2s^2\,2p^2$ per C) are in a planar $\sigma$-orbital (with $sp^2$ hybridization). The remaining two are in a $\pi$-orbital orthogonal to the surface of the tube~\cite{jap}. These electrons are confined into a layer of $1\div 2$~\AA~around the geometric surface of the nanotube.    

We estimate that a $10$~nm nanotube is $\approx 96\%$ empty. We must however take into account the incoherent scatterings between the channeled nuclei and the atoms of the wall. When a nucleus gets close to the boundaries of the nanotube it scatters off electrons and nuclei  as it would be doing in a amorphous medium but, in the nanotube  case,  the density of electrons and nuclei depends on the position of the channeled particle.

For a numerical study of dechanneling we simulate the motion of a channeled ion being perturbed by Coulomb scatterings with mean square diffusion angles $\delta$ depending on the position in the nanotube
\begin{equation}
\langle \delta^2\rangle = 16 \pi Z^2 e^4 \frac{\Delta z}{T^2} \left(n_n(r) Z^2 L_n + n_e(r) L_e\right)
\end{equation}
$\Delta z$ being the distance traversed in the longitudinal direction, $T$ the total kinetic energy and $L_n = \ln(191\, Z^{-1/3})$, $L_e = \ln(1194\, Z^{-2/3})$ (see {\it e.g.}~\cite{russi}). 
The scattering produces a change in the direction of the particle, redistributing longitudinal and transverse kinetic energy. This causes a change in the total transverse energy, which is no longer strictly conserved as required for channeling to be maintained. The transverse energy of the particle might even increase above the confining potential barrier.

Studying numerically the classical trajectories of ions in nanonotubes, we find that nuclei with smaller angular momenta penetrate closer to the boundaries and, because of this, dechannel quickly whereas nuclei with larger angular momenta dechannel rarely. We conservatively find that about $25\%$ of the nuclei dechannel before traversing 200~$\mu$m.

\item {\bf \emph{Interstices.}} A carbon ion scattered in the interstices among aligned carbon nanotubes is also experiencing a repulsive potential when approaching the nanotube surfaces from the outside. Therefore  interstices play the role of channels as well as the interior of  single CNTs do. The problem is mathematically equivalent to a billiard problem (Sinai problem) and the numerical results of this analysis (based on some exact solutions) will be presented in a separate paper.  The quantity of interest is the average time for the transverse motion to reach the boundaries of the array (starting from a random point within the bulk) when compared with the time needed to span the length of CNT's ($\approx 200~\mu$m) given a certain initial longitudinal kinetic energy $E_\parallel$. 

\item {\bf \emph{Potential pattern at boundaries.}} We are able to study a multiple pattern structure of nanotubes with regions populated by CNTs with constant potential barriers at the surface mixed with regions with variable potentials. The collective directional properties of the CNT arrays are not found to be qualitatively altered.  

\item {\bf \emph{Secondary channelings.}}  When the channeling initial conditions are {\it not} met, most likely when the carbon nanotube array is oriented orthogonal to the Cygnus direction, carbon ions undergo multiple scatterings in the CNT array. These are C-C scatterings which  mostly occur in the forward direction (the $\theta=\pi/2$ case has zero momentum -- when  generating the scattering angles $\eta$ in the laboratory frame we use the probability distribution $\propto \cos \eta/\sin^3\eta$ which is the one for equal masses of target and projectile). We find that when the initial angle $\aleph$ with respect to the plane orthogonal to CNTs axes is small, the C will eventually be absorbed in the bulk. 

However, increasing the value of the initial $\aleph$ angle,  the scattered ion could match the channeling conditions as a consequence of one of the multiple collisions with C nuclei.

If  $\aleph$ is sufficiently large, it might take a few collisions for this matching to occur. 
This is a crucial point as there is a finite energy loss per collision and, for the C ion to be channeled up to the open end of the nanotube, it must have sufficient longitudinal kinetic energy.

This is what we call a {\it secondary channeling}: see the results in Fig.~\ref{altra2}.  The overall effect of secondary channeling is that of relaxing the strict initial conditions on $\theta$ allowing larger initial values which, however, as can be read in Fig.~\ref{altra2}, cannot spoil the anisotropic response we want to achieve.

\end{enumerate}

We stress here that $E_\parallel$ can be as low as 1~keV differently from channeling in other materials where energy loss due to electrons may  effectively stop low energy ions. This condition is particularly important in the search of $\approx 10$~GeV mass WIMPS which, assuming the maximum energy transfer, might scatter carbon nuclei with $T\lesssim 10$~keV.
   
 For the sake of illustration assume $\approx$ 10$^{14}$, $10$~nm diameter CNTs on a 10$\times$10~cm$^2$ substrate. This might be obtained from  CNT ropes suitably cut and oriented.  Since the surface density of a graphene sheet is 1/1315 g/m$^2$~\cite{massa}  a single-wall CNT weights about $50 \times10^{-16}$~grams, therefore setting nanotubes on both faces of the substrate one reaches $\approx 1$ grams on a single substrate or $\approx$10 Kg on 100 thin stacked panels $1\times 1$~m$^2$ each. 

 We also  remark that in this detection scheme most of the mass is concentrated into the ``target" CNT: being the density of 100 mbar Ar about 2 10$^{-4}$ g/cm$^3$  the passive (gas)  to active (CNT)  mass ratio would 10$^{-3}$. The whole system can be installed on a movable system pointing continuously towards the Cygnus.

  A detailed study of the ion detection system and of backgrounds is the subject of a future experimental program. In particular we plan to expose  the detector module described above to a beam of few keV carbon ions: this would help to verify the channeling properties  of CNTs  with a carbon ion beam of known direction allowing  to test the model  outlined in this paper.  Moreover,  an intense beam of collimated neutrons with tens of keV kinetic energy would allow us to mimic the WIMP-carbon elastic scattering and validate the   directional property of the detector we propose. This  type of beam, with a wide range of neutron energies, is available at the nTOF experimental area at CERN.
  
 \section{Conclusions} 
 We presented a new  concept of detection for directional dark matter searches and analyzed its potentialities through a numerical study of its fundamental unit: a  single-wall carbon nanotube having a variable orientation with respect to the WIMP wind. We also  find that an aligned array of CNTs is collectively cooperating at channeling ions with low transverse kinetic energy along the extension of nanotubes. The conditions for channeling to occur are in the initial kinematic configuration of the ion scattered off by the WIMP. This in turn depends on whether or not the CNT array is directed in the Cygnus direction.  
  
Once multiple interactions are taken into account, due to secondary channeling, the carbon nanotube array detector offers an improved acceptance and  signal strength.  The almost bidimensional structure of CNT's bodies guarantees that if any WIMP-carbon nucleus interaction takes place, the scattered carbon nucleus will either undergo a primary or secondary channeling or just be stopped within the CNT array.  The kinetic energy of the recoiling nucleus, to be reconstructed by the apparatus, can be correlated to the channeling probability and the mass of the WIMP  can be estimated by a comparison with the expected counts.   

We believe that these results could motivate an experimental program which we intend to pursue in the near future.

\section*{Appendix~1: WIMP-Nucleus Collisions}
\begin{figure}[htb!]
 \begin{center}
   \includegraphics[width=6truecm]{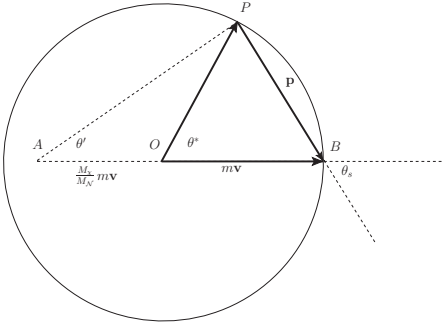}
 \end{center}
\caption{\small We are considering the non-relativistic elastic scattering of a DM particle $\chi$ impinging on a nucleus $\n$ initially at rest. The, non-relativistic, initial momentum of $\chi$ in the centre of mass frame (COM) is therefore $m\bm v (\equiv M_\chi v_{\chi}^{\mathrm{_{COM}}})$, with $m=(M_\chi M_\n)/(M_\chi+M_\n)$ being the reduced mass. After the elastic scattering has taken place, because of the   conservation of   kinetic energy, the initial incoming momentum $OB$ is rotated by $\theta^*$ (the COM scattering angle) into $OP$, $P$ being a generic point on the circle -- the position of $P$ fully characterises the scattering process.   As is readily verified, $AB=M_\chi\bm v$ coincides with the incoming momentum of $\chi$ in the laboratory frame (LAB). Depending on $M\chi \gtrless M_\n$ the point $A$ will lie outside/inside the disk of radius $mv$.  On the other hand, the momentum of $\chi$ after the scattering, in the LAB frame, is obtained by the  transformation $\bm k^\prime=m\bm v^\prime +M_\chi \bm V$ where $m\bm v^\prime$ is the rotated vector $OP$ and $\bm V$ is the velocity of the COM, $\bm V=M_\chi v/(M_\chi+M_\n)$. Thus $\bm k^\prime=m\bm v^\prime +(M_\chi/M_\n)  m\bm v=AP$.  Moreover, $PB=AP-AB=\bm k^\prime-\bm k=\bm p$, $\bm p$ being the momentum of the scattered nucleus in the LAB frame. The $\theta_s$($<\pi/2$) angle is found between the initial LAB direction of $\chi$ and the $\n$ recoil momentum in the LAB.    \label{circle}}
\end{figure}

The velocity $\bm v(\lambda)$ in Fig.~\ref{frame} is given by
\be
\bm v(\lambda)=\bm v_0 \cos\lambda+\bm n(\bm n\cdot \bm v_0 )(1-\cos\lambda)+(\bm v_0\times\bm n)\sin\lambda
\label{vrot}
\ee
where we let $\bm v_0=\bm v(\lambda=0)$ be that position of $\bm v(\lambda)$ in which $\varphi_\chi=\varphi$ and $\theta=\theta_\chi+\theta_s$ with $\theta_s<\pi/2$ (see Fig.~\ref{circle}). Then we let $\lambda$ vary  in the interval $\lambda\in [0,2\pi]$~\footnote{Let us check some cases. Rotating $\bm n$ at fixed $\varphi$ in such a way that $\bm v_0$ has  $\varphi_\chi=\varphi+\pi$ and $\theta_\chi=\theta_s-\theta$ produces the same expression for $\bm v(\lambda)$. Rotating $\bm n$ at fixed $\varphi$ until $\bm v_0$ is just along $\hat z$, we have  $\varphi=\varphi_\chi$ when $\lambda\to\lambda+\pi$; in this point $\theta=\theta_\chi-\theta_s$. This produces the same expression of $\bm v(\lambda)$ as given above (it is essential that $\theta_s<\pi/2$). When $\bm v_0$ is in the direction with $\theta_\chi=0$, fixing $\varphi_\chi=\varphi$ is not essential as $\bm v_0$  has no $\varphi_\chi$ dependency left. Similarly when $\theta=0$ there is no need to fix $\varphi_\chi=\varphi$. }.

We will eventually be interested in the line integral along $\c$ of some velocity distribution $f(|\bm v|)$ as given by
\be
\int_{\c}f ds=\int_0^{2\pi}f(|\bm v(\lambda)|)\left| \frac{d}{d\lambda} \bm v(\lambda)\right| d\lambda=v\sin\theta_s\int_0^{2\pi}f(|\bm v(\lambda)|)d \lambda
\label{curvilineo}
\ee
The factor $\sin\theta_s$ found in the curvilinear integration, deriving $\bm v(\lambda)$ with respect to $\lambda$, is indeed part of the $d^3v$ velocity element which can be read by Fig.~\ref{frame} to be
\be
d^3v =(v\sin\theta_sd\lambda)(vd\theta_s)dv=(v^2dv)\, d\cos\theta_s \,d\lambda
\ee
The $\lambda$ integration is carried out at a {\it  fixed} $\theta_s$, which means, as we will see below,  that the nucleus recoiling kinetic energy $T$ and incident velocity $v$ are fixed. 

With reference to Fig.~\ref{circle} we can  determine the expression for $\cos\theta_s$ as a function of $T$ (the recoiling nucleus energy), $v$ and the mass  $\mu$ defined below. The scattering $\chi+\n\to\chi+\n$ is elastic and non-relativistic. The momentum $\bm p$ of the recoiling  $\n$ is given by (see Fig.~\ref{circle})
\be
2m v\sin\frac{\theta^*}{2}=|\bm p|
\ee 
where $\theta^*$ is the scattering angle in the center of mass and $m$ the reduced mass of the system, so that 
\be
2M_\n T=4m^2v^2\sin^2\frac{\theta^*}{2}
\ee
or simply, since $2\theta_s+\theta^*=\pi$, where $T$ is the recoiling nucleus {\it kinetic} energy
\be
\frac{2T}{\mu v^2}=\sin^2\frac{\theta^*}{2}=\sin^2\left(\frac{\pi}{2}-\theta_s\right)=\cos^2\theta_s
\label{cinet}
\ee
where 
\be
\mu\equiv \frac{4M_\chi^2 M_\n}{(M_\chi+M_\n)^2}
\label{muu}
\ee
Thus one obtains
\be
\cos\theta_s=\sqrt{\frac{2T}{\mu v^2}}
\label{coseno}
\ee
whereas $v_{\rm min}=\sqrt{2T/\mu}$. 


If $n_\chi$ is the incident dark matter (DM) particle (number) density, we define an initial-velocity-averaged  rate distributed in the solid angle $d\Omega_{\bm n}$ relative to the recoiling nucleus direction as in Fig.~\ref{frame}
\be
\frac{d\Gamma}{d\Omega_{\bm n}}:=n_\chi \left\langle v\,\frac{d\sigma}{d\Omega_{\bm n}}\right\rangle_{\bm v}
\label{definizione}
\ee
Using the standard formalism, the cross section distribution with respect to the solid angle $d\Omega_{\bm n}$ where the nucleus recoils with energy $E$ is, in the laboratory frame (and in the non-relativistic limit for $v\approx 10^{-3} c$)
\bea
\frac{d\sigma}{d\Omega_{\bm n}}&=&\frac{1}{32\pi^2  M_\n |\bm k|}\int dE \, |\bm p^\prime|\,  \int\frac{d^3 k^\prime}{2\en^\prime} \delta^4(k-q-k^\prime)\, \overline{\sum_{\rm pol}}|M|^2=\notag\\
&\equiv&\frac{1}{32\pi^2  M_\n M_\chi v}  \int dE\,  |\bm p^\prime|\, \int\frac{d^3 k^\prime}{2\en^\prime} \delta^4(k-q-k^\prime)\, \int\, d\en^\prime\, 2\en^\prime\theta(\en^\prime)\delta(k^{\prime 2}-M_\chi^2)\overline{\sum_{\rm pol}}|M|^2=\notag\\
&=&\frac{1}{32\pi^2  M_\n M_\chi v}  \int dE\,  |\bm p^\prime|\, \delta(q^2-2k\cdot q)\,G^2_A |F_A(\bm q^2)|^2 \, \Lambda^4=\notag\\
&\simeq&\frac{1}{32\pi^2 M_\n M_\chi v}  \frac{\sqrt{2M_\n T}}{(M_\n+M_\chi)}G^2_A |F_A(2M_\n T)|^2\, \Lambda^4
\label{xsect}
\eea
where $(\en,\bm k)$ and $(\en^\prime,\bm k^\prime)$ are related to the incident and scattered DM particle respectively \footnote{Here we might cancel $M_\n\, v$ against $\sqrt{2M_\n T}$. $v$ will  cancel out when evaluating the average $\langle n_\chi \,v\, d\sigma/d\Omega_{\bm n}\rangle_{\bm v}$. }. $G_A$ is some unknown effective DM-nucleus interaction coupling whereas $F_A(\bm q^2)$ is a nuclear form factor (adimensional) and $\Lambda$ has dimensions of GeV -- both depend on the $A$ nucleus involved. 
Here $q$ is the transferred momentum $q=k-k^\prime=p^\prime-p$, $p$ and $p^\prime$ being the initial and final momenta of the nucleus $p=(M_\n,\bm 0)$, $p^\prime=(E,\bm p^\prime)$. Indeed $\delta(q^2-2k\cdot q)=\delta(f(E))$ where 
\be
f(E)=2M_\n(M_\n+\en)-2E(M_\n+\en)+2|\bm k||\bm p^\prime|\cos\theta_s
\ee
and $|\bm p^\prime|=\sqrt{E^2-M_\n^2}$. 
The $f(E)=0$ equation has two roots. One is $E_1=M_\n$, but $1/|f^\prime(E)|_{E_1}=0$. The other one is
\be
E_2=M_\n\frac{(\en+M_\n)^2+k^2\cos^2\theta_s}{(\en+M_\n)^2-k^2\cos^2\theta_s}
\ee
where $k\simeq M_\chi v$, $\en\simeq M_\chi$ (in the non-relativistic limit) and, using~(\ref{muu}) and~(\ref{coseno}), we have  $E_2=M_\n(2M_\n+T)/(2M_\n-T)$ and $1/|f^\prime(E)|_{E_2}=1/(M_\n+M_\chi)$. Thus $F(\bm q^2=-2M_\n^2+2M_\n E)_{E=E2}\simeq F(2M_\n T)$~\footnote{$(-2M_\n^2+2M_\n E)_{E=E_2}=4M_\n^2 T/(2 M_\n-T)$, where we can neglect $T$ with respect to $M_\n$ in the denominator.},
since for low momentum transfers with respect to the mass $M_\n$,  $q^2\approx -\bm q^2$.

Therefore we find
\be
\frac{d\Gamma}{d\Omega_{\bm n}}=\frac{n_\chi G^2_A\,\Lambda^4\, F_A(2M_\n T)^2 \sqrt{2M_\n T}}{32\pi^2  M_\chi M_\n (M_\n+M_\chi)}\int d^3v \,f(\bm v+\bm w) 
\label{rate}
\ee
When adding $\bm w$ to $\bm v$ in the frame of Fig.~\ref{frame} one gets $\bm v+\bm w(t)=\bm v^\prime$, the latter being the velocity of DM particles in the Galactic frame, where they are supposed to obey a Mawell-Boltzmann velocity distribution $f(\bm v^\prime)$.  

The scattering amplitude does not depend on the scattering plane which is spanned by $\bm v(\lambda)$ and $\bm n$ in Fig.~\ref{frame}. Therefore we can choose $\lambda=0$, place the vector $\bm k^\prime$ in that plane (with  $\theta^\prime$ being the  angle formed by  $\bm k^\prime$ and $\bm k=M_\chi \bm v_0$ -- see $\theta^\prime$ in Fig.~\ref{circle}) and use the distribution $d\sigma/d\Omega_{\bm n}$ computed. 

We are interested in a distribution in the nuclear recoil kinetic energy variable $T$, not in $d\cos\theta_s$, which is not measurable. However we know from~(\ref{cinet}) that 
\be
T=\frac{1}{2}\mu v^2\cos^2\theta_s
\ee
therefore to change variables we consider 
\bea
&&\int d^3 v\, f(\bm v+\bm w) \, (\cdots)=\int v^2 dv\, d\cos\theta_s \, d\lambda \, f(\bm v(\lambda)+\bm w) \, (\cdots) \int\,dT\,\delta\left( T-\frac{1}{2}\mu v^2\cos^2\theta_s \right)=\notag\\
&&=\int dT \int_{v_{\rm min}}^{v_{\rm max}} v^2\, dv\, \int_0^{2\pi} d\lambda \int d\cos\theta_s \, f(\bm v(\lambda)+\bm w) (\cdots)\, \delta\left( T-\frac{1}{2}\mu v^2\cos^2\theta_s \right)=\notag\\
&& =\int dT \int_{v_{\rm min}}^{v_{\rm max}} v^2\, dv\, \int_0^{2\pi} d\lambda  \, f(\bm v(\lambda)+\bm w) (\cdots)\, \frac{1}{v\sqrt{2\mu T}}
\eea
where in the last line we carried out the integration in $d\cos\theta_s$ and it is meant that in the integrand function we have to substitute $\cos\theta_s$  by $\sqrt{2T/\mu v^2}$. By $(\cdots)$ we mean the explicit result obtained from $v\,d\sigma/\Omega_{\bm n}$. Since eventually we will consider rate distributions in the $T$ variable, we are left with 
\be
\int_{v_{\rm min}}^{v_{\rm max}} v\, dv\, \int_0^{2\pi} d\lambda  \, f(\bm v(\lambda)+\bm w) (\cdots)\, \frac{1}{\sqrt{2\mu T}}
\ee

Therefore, returning to~(\ref{rate}),  and from $d\Omega_{\bm n}=d\cos\theta \, d\varphi$ we can write 
\be
\frac{d\Gamma}{dT\, d\cos\theta}=\frac{n_\chi G^2_A\,\Lambda^4\, F_A(2M_\n T)^2}{64\pi^2 M_\chi^2  M_\n}\int_{\sqrt{2T/\mu}}^{v_{\rm max}} dv\,v \int_0^{2\pi} d\varphi \,  \int_0^{2\pi} d\lambda\, f(\bm v(\lambda)+\bm w)
\ee 
where $v_{\rm max}=232+550$~km/sec. 

The velocity distribution is defined by
\be
f(\bm v+\bm w)=\beta e^{-\alpha (\bm v+\bm w)^2}
\ee
where $\alpha\sim 1/v_0^2,\beta\sim 1/v_0^3$ ($v_0$ is usually taken to be $v_0\simeq235$~km/sec), and the distribution peaks at $\bm v=-\bm w$, {\it i.e.} along the wind of $\chi$ particles, see Eq.~(\ref{wut}).
Therefore we have
\be
f(\bm v+\bm w)= \beta\,e^{-\alpha(A(v,T,\theta,t)+B(v,T,\theta,t)\cos\lambda)}
\ee
where, using Eq.~(\ref{vrot}), we obtain~(\ref{uno}) and~(\ref{due})
using the formula for $\bm w$ in~(\ref{wut}). Observe that there is no dependency left on the  $\varphi$ angle.

Incidentally, we note that a reduction in terms of the  modified Bessel function of the first kind is possible 
\be
\int_0^{2\pi}d\lambda\, e^{-\alpha(A+B\cos\lambda)}=2\pi e^{-\alpha\, A}\, I_0(\alpha\, B)
\ee
so that Eq.~(\ref{rate2}) is obtained~\footnote{We use $M_\chi,M_\n=\hash$~GeV,  $T=\hash$~keV, $n_\chi=\hash$~cm$^{-3}$. We wish to express the integral in a adimensional form. This requires a factor of $1/c$ in front of the integral. In order to express $d\Gamma/dT $ in units of 1/(keV$\cdot$ sec) we have to include an overall constant of  $K_1=1.16\times 10^{-23}$ on the rhs of~(\ref{rate2}) -- $K_1$ accounts also for  the conversion of cm$^{-3}$ to GeV$^3$. $d\Gamma/dT$ has also to be multiplied by the number of target nuclei per kilogram $N_T=N_A/MW$, where $MW=$ molecular weight of the detector.  Therefore, using a molecular weight of graphene of $\approx 12$~gr/mole to express $d\Gamma/dT$ in {\it (counts per day)}/kg/keV we have to include a constant overall factor of $K=24\times 60^2\times K_1K_2=5\times 10^{7}$ where  $K_2=5\times 10^{25}$~kg$^{-1}$. \label{fook}}.
The non-relativistic kinematics of the DM-nucleus scattering involves the factorization of the nuclear form factor so that the angular distribution is universal, depending only on the Maxwell-Boltzmann assumption of the DM velocity distribution.  


In the following we discuss the $G_A$ factor appearing in formula~(\ref{rate2}). We will take it from DAMA data on WIMP-Na scattering at $M_\chi\simeq 11~\gev$. 
Indeed this is the situation closer to the case discussed in this paper.

Here by $\sigma_{\chi\n}$ the following expression is meant
\be
\sigma_{\chi\n}=\int_{0}^{4\mu_{\chi\n}v^2} d|\bm q|^2\, \frac{d\sigma (\bm q=0)}{d|\bm q|^2} 
\label{fcp}
\ee
where $|\bm q|$ is the momentum transfer. 
With a calculation analogous to~(\ref{xsect}) we compute the distribution \mbox{$d\sigma(|\bm q|=0)/d|\bm q|^2$} and from~(\ref{fcp}) we get 
\be
\sigma_{\chi \n}= \frac{1}{16\pi (M_\chi+M_\n)^2}\, G_A^2\, \Lambda^4
\label{noic}
\ee
On the other hand from DAMA data on WIMP-Na collisions, it is found that
\be
\sigma_{\chi p}=2\times 10^{-4}~{\rm pb}=5.15\times 10^{-13}~\mathrm{GeV}^{-2}
\label{sigmachip}
\ee
If we assume $\Lambda\sim \sqrt{s}\sim M_\chi+M_p\simeq M_\chi$  then from~(\ref{noic}) we can write
\be
\sigma_{\chi p}= \frac{G^2\, M_\chi^4}{16\pi (M_\chi+M_p)^2}\simeq \frac{G^2M_\chi^2}{16\pi}
\label{noic2}
\ee
so that 
\be
G=4.6\times 10^{-7}~\mathrm{GeV}^{-2}
\ee
Since we know that  $\sigma_{\chi \n}/\sigma_{\chi p}\simeq M_\chi^2/(M_\chi+M_\n)^2 \, A^4 $,  
from (\ref{noic}) (with $\Lambda\sim \sqrt{s}\simeq M_\chi+M_\n$) and (\ref{noic2}) we have
\be
G_A=\frac{A^2}{4} G= 1.6\times 10^{-5}~\mathrm{GeV}^{-2}
\ee
for carbon ($A=12$ and $M_\chi=M_{\n=C}$). 

As a cross check of the method for computing  $G_A$, we use~(\ref{rate2}) to confront directly with DAMA data 
on modulation (time dependency of $d\Gamma/dT$) which are generally used to fix  $\sigma_{\chi p}$ in~(\ref{sigmachip}). 
To make this comparison we have to rescale $G_A$ to the value predicted for sodium, and, following the discussion 
in footnote~\ref{fook}, change $K$ in~(\ref{rate2}) to be related to the molecular weight $MW$ of the NaI DAMA detector. 
Rescaling
\be
G_{A^\prime}=\left(\frac{A^\prime}{A}\right)^2 \, G_A
\ee
we expect
\be
G_{\rm Na}= 5.8\times 10^{-5}~\gev^{-2}
\label{befico}
\ee

Confronting with DAMA modulation data, using a value of $M_\chi\approx 11~\gev$ and rescaling the $K$ factor by $12/150$, as can be understood from footnote~\ref{fook},
the best fit walue obtained is $G_{\rm Na}\simeq 5.3\times 10^{-5}~\gev^{-2}$ which is perfectly consistent with the one computed in~(\ref{befico}) and thus with the $\sigma_{\chi p}$ 
value in~(\ref{sigmachip}).


\section*{Acknowledgements}
We wish to thank M.~Diemoz for having stimulated our interest on this topics and P~Lipari for a thorough introduction on the subject of directionality and for his collaboration in the early stages of this work. We also thank C. Savage for a useful response in answer to our queries. We thank some members of the group of condensed matter physics in Rome,  C.~Mariani, M.G.~Betti and A.~Polimeni for informative discussions on carbon nanotubes and nanowires and S.~Orlanducci and M.L.~Terranova for providing information on the synthesis of CNTs. A TPC-GEM was made available by F.~Murtas.   G.C. acknowledges partial support from ERC Ideas Consolidator Grant CRYSBEAM  G.A. n.615089.

\bibliographystyle{unsrt}

\begin{thebibliography}{99}
\bibitem{dali}  
 R.~Bernabei, P.~Belli, F.~Cappella, V.~Caracciolo, S.~Castellano, R.~Cerulli, C.~J.~Dai and A.~dÕAngelo {\it et al.},
  Phys.\ Part.\ Nucl.\  {\bf 46}, no. 2, 138 (2015).


\bibitem{cogent} 
  C.~E.~Aalseth, P.~S.~Barbeau, J.~Colaresi, J.~I.~Collar, J.~Diaz Leon,
  J.~E.~Fast, N.~Fields and T.~W.~Hossbach {\it et al.},
  Phys.\ Rev.\ Lett.\  {\bf 107}, 141301 (2011)
  [arXiv:1106.0650 [astro-ph.CO]].



\bibitem{cdms} 
  R.~Agnese {\it et al.}  [CDMS Collaboration],
  Phys.\ Rev.\ Lett.\  {\bf 111}, no. 25, 251301 (2013)
  [arXiv:1304.4279 [hep-ex]].
  
\bibitem{cresst}
  G.~Angloher, M.~Bauer, I.~Bavykina, A.~Bento, C.~Bucci, C.~Ciemniak,
  G.~Deuter and F.~von Feilitzsch {\it et al.},
  Eur.\ Phys.\ J.\ C {\bf 72}, 1971 (2012)
  [arXiv:1109.0702 [astro-ph.CO]];
   G.~Angloher {\it et al.}  [CRESST-II Collaboration],
  Eur.\ Phys.\ J.\ C {\bf 74}, no. 12, 3184 (2014)
  [arXiv:1407.3146 [astro-ph.CO]].
  
 
 

\bibitem{Akerib:2013tjd}
 D.~S.~Akerib {\it et al.}  [LUX Collaboration],
 Phys.\ Rev.\ Lett.\  {\bf 112} (2014) 9,  091303
 [arXiv:1310.8214 [astro-ph.CO]];
 Z.~Ahmed {\it et al.}  [CDMS-II Collaboration],
  Science {\bf 327}, 1619 (2010)
  [arXiv:0912.3592 [astro-ph.CO]];
  Z.~Ahmed {\it et al.}  [CDMS-II Collaboration],
  Phys.\ Rev.\ Lett.\  {\bf 106}, 131302 (2011)
  [arXiv:1011.2482 [astro-ph.CO]].


 
    \bibitem{spergel} 
  D.~N.~Spergel,
  Phys.\ Rev.\ D {\bf 37}, 1353 (1988);
  
  
  A.~Drukier and L.~Stodolsky,
  Phys.\ Rev.\ D {\bf 30}, 2295 (1984);

  M.~W.~Goodman and E.~Witten,
  Phys.\ Rev.\  D {\bf 31}, 3059 (1985);

  A.~K.~Drukier, K.~Freese and D.~N.~Spergel,
  Phys.\ Rev.\ D {\bf 33}, 3495 (1986);

  F.~T.~Avignone, III, R.~J.~Creswick and S.~Nussinov,
  arXiv:0807.3758 [hep-ph].

\bibitem{damaele} 
  R.~Bernabei, P.~Belli, F.~Montecchia, F.~Nozzoli, F.~Cappella, A.~Incicchitti, D.~Prosperi and R.~Cerulli {\it et al.},
Phys.\ Rev.\ D {\bf 77}, 023506 (2008)
[arXiv:0712.0562 [astro-ph]].
\bibitem{drift} See talk by Dinesh Loomba in the 42nd SLAC Summer Institute, {\it Shining Light on Dark Matter},~{\tt https://indico.cern.ch/event/297618/other-view?view=standard} and S. Burgos et al., Astropart.\ Phys.\  {\bf 31}  261-266 (2009).

\bibitem{nuclemul}
  A.~Alexandrov, T.~Asada, N.~D'Ambrosio, G.~D.~Lellis, A.~D.~Crescenzo, N.~D.~Marco, S.~Furuya and V.~Gentile {\it et al.},
  JINST {\bf 9}, no. 12, C12053 (2014).

\bibitem{Drukier:2012hj} 
  A.~Drukier, K.~Freese, D.~Spergel, C.~Cantor, G.~Church and T.~Sano,
  arXiv:1206.6809 [astro-ph.IM].


\bibitem{lis}
 K.~Freese, M.~Lisanti and C.~Savage,
  Rev.\ Mod.\ Phys.\  {\bf 85}, 1561 (2013)
  [arXiv:1209.3339 [astro-ph.CO]].
 
 \bibitem{russi} N.K. Zhevago and V.I. Glebov, Journal of Experimental and Theoretical Physics, Vol. 91, No. 3 (2000), pp. 504-514; 
  see also X.~Artru, S.~P.~Fomin, N.~F.~Shulga, K.~A.~Ispirian and N.~K.~Zhevago,
  Phys.\ Rept.\  {\bf 412}, 89 (2005).
  
  \bibitem{lind}
J.~Lindhard,
Kongel.\ Dan.\ Vidensk.\ Selsk.\, Mat.-Fys.\ Medd. \ {\bf 34}, No. 14 (1965). 

  \bibitem{gelmi}  G.~B.~Gelmini,
  J.\ Phys.\ Conf.\ Ser.\  {\bf 384}, 012007 (2012)
  [arXiv:1201.4560 [astro-ph.CO]].

\bibitem{jap}
T. Ando, J. Phys. Soc. Jpn. 74, pp. 777-817 (2005); See also the review 
NPG Asia Materials (2009) 1, 17Ð21; 


  \bibitem{massa} C.~Laurent and E.~Flahaut and A.~Peigney,  Carbon, {\bf 48} n.10  2994 (2010).
  
  
\bibitem{Burgos:2007tt}
  S.~Burgos {\it et al.},
  Nucl.\ Instrum.\ Meth.\ A {\bf 584} (2008) 114
  [arXiv:0707.1758 [physics.ins-det]].
  
  \bibitem{Jungman} 
  G.~Jungman, M.~Kamionkowski and K.~Griest,
  Phys.\ Rept.\  {\bf 267}, 195 (1996)
  [hep-ph/9506380].
  
  
\end{thebibliography}

\end{document}